\documentclass[letterpaper, 10 pt, conference]{ieeeconf}  

\usepackage{graphicx}
\usepackage{tabularx,booktabs}
\usepackage{color}
\usepackage{multicol}
\usepackage{amsmath,amssymb}
\usepackage{amsfonts}
\usepackage{textcomp}
\usepackage{psfrag}
\usepackage{multimedia}
\usepackage{fancybox}
\usepackage{algorithm}
\usepackage{algorithmic}
\usepackage{varioref}
\usepackage{import}
\usepackage{framed,xcolor}
\usepackage{color}
\usepackage{float}
\usepackage{etoolbox}
\usepackage{soul}

\IEEEoverridecommandlockouts                              

\definecolor{seb}{rgb}{0.8,1,0.8}

\definecolor{hos}{rgb}{0.91,0.84,0.42}

\newcommand{\vect}[1]{\ensuremath{\boldsymbol{\mathrm{#1}}}}

\title{\LARGE \bf
Approximate Robust NMPC using Reinforcement Learning
}

\author{Hossein Nejatbakhsh Esfahani, Arash Bahari Kordabad, S\'ebastien Gros
\thanks{Authors are with Department of Engineering Cybernetics, Norwegian
	University of Science and Technology (NTNU), Trondheim, Norway.
        {\tt\small \{hossein.n.esfahani, Arash.b.kordabad, sebastien.gros\}@ntnu.no}}%
}

\begin{document}
\maketitle
\begin{abstract}

We present a Reinforcement Learning-based Robust Nonlinear Model Predictive Control (RL-RNMPC) framework for controlling nonlinear systems in the presence of disturbances and uncertainties. An approximate Robust Nonlinear Model Predictive Control (RNMPC) of low computational complexity is used in which the state trajectory uncertainty is modelled via ellipsoids. Reinforcement Learning is then used in order to handle the ellipsoidal approximation and improve the closed-loop performance of the scheme by adjusting the MPC parameters generating the ellipsoids. The approach is tested on a simulated  Wheeled Mobile Robot (WMR) tracking a desired trajectory while avoiding static obstacles.
\end{abstract}

\section{INTRODUCTION}

\par 
Nonlinear Model Predictive Control (NMPC) is an optimization based control approach operating in a receding horizon \cite{rawling}, which is often adopted for its ability to handle linear/nonlinear state and input constraints. MPC offers rigorous theoretical properties (such as recursive feasibility, constraint satisfaction and stability), assuming that an accurate model of the plant is available. There are many autonomous systems (such as ground vehicles, marine robots and unmanned aerial vehicles) for which NMPC-based algorithms have been adopted \cite{uav}, \cite{agv}, \cite{usv}.
\par 
Reinforcement Learning (RL) is a powerful tool for tackling Markov Decision Processes (MDP) without depending on a model of the probability distributions underlying the state transitions of the real system \cite{sutton}. Indeed, most RL methods rely purely on observed state transitions, and realizations of the stage cost in order to increase the performance of the control policy.
\par 
Recently, the integration of machine learning in MPC has been invesstigated, with the aim of learning the model of the system, the cost function or even the control law directly \cite{sergio1,sergio2}. For computational reasons, simple models are usually preferred in the MPC scheme. Hence, the MPC model often does not have the structure required to correctly capture the real system dynamics and stochasticity. As a result, MPC delivers a reasonable but suboptimal approximation of the optimal policy. Choosing the MPC parameters that maximises the closed-loop performance for the selected MPC formulation is a difficult problem. Indeed, e.g. selecting the MPC model parameters that best fit the model to the real system is not guaranteed to yield the best closed-loop performance that the MPC scheme can achieve \cite{gros2019data}. In \cite{gros2019data,gros2020reinforcement}, it is shown that adjusting the MPC model, cost and constraints can be beneficial to achieve the best closed-loop performances, and Reinforcement Learning is proposed as a possible approach to perform that adjustment in practice. Further recent research have focused on MPC-based policy approximation for RL~\cite{koller2018learning,mannucci2017safe,zanon2019practical,bahari2021reinforcement,nejatbakhsh2021reinforcement}, where it is shown that a single Model Predictive Control (MPC) scheme can capture the optimal value function, action-value function, and policy of an MDP, even if the MPC model is inaccurate, hence providing a valid and generic function approximator for RL.

\par 
Model-plant mismatch and disturbances can be treated via Robust NMPC (RNMPC) techniques. For linear MPC models and polytopic disturbance models and constraints, tube-based MPC techniques provides computationally effective techniques \cite{Wang2019}, \cite{RMPC6}. Treating nonlinear MPC models or generic disturbances and constraints is more challenging \cite{RMPC1}.
Researchers in \cite{RMPC2} proposed to use a tube-based MPC with a Min-Max differential inequality. Multi-stage or Scenario-tree NMPC scheme was proposed in \cite{RMPC3,RMPC4} as a real-time NMPC that accounts for the uncertain influence and generates decisions to control a nonlinear plant in a robust sense. These approaches remain challenging for problems that are not of small scale.

\par
In this paper we model the propagation of perturbations in the state dynamics via ellipsoids, based on the linearization of the system dynamics and constraints on the nominal trajectories and using a Gaussian disturbance model. We propose to adjust this scheme using the RL method in order to tailor this inaccurate uncertainty model to the real system and achieve a better closed-loop performance. A fast convergence of the adjustable parameters of RNMPC is achieved via a second-order Least Square Temporal Difference Q-learning (LSTDQ).
\par 
This paper is organised as follows. In Section \ref{sec:1}, the proposed approximate robust NMPC is formulated. The combination of the RL algorithm and RNMPC is detailed in Section \ref{sec:2}. A  mobile robot under uncertainties is adopted in Section \ref{sec:3} to be controlled by the proposed RL-RNMPC for a trajectory tracking scenario associated to elliptic static obstacle avoidance in presence of model uncertainty. Finally, conclusions and future work are given in Section \ref{sec:4}.

\section{Ellipsoidal-based Robust NMPC}\label{sec:1}

We consider a model based on the nonlinear dynamics
\begin{align}
	\label{eq:Model}
	\vect x_+ = \vect f\left(\vect x,\vect u,\vect d\right)
\end{align}
where $\vect d$ is a stochastic variables affecting the state evolution. We assume that the real system is unknown and imperfectly represented by \eqref{eq:Model}. A nominal NMPC scheme based on \eqref{eq:Model} reads e.g. as:
\begin{subequations}
\label{eq:ClassicMPC}
	\begin{align}
		\min_{\bar{\vect x},\bar{\vect u}}&\quad T\left(\bar{\vect x}_N\right) + \sum_{k=0}^{N-1} L\left(\bar{\vect x}_k,\bar{\vect u}_k\right) \\
		\mathrm{s.t.}&\quad \bar{\vect x}_{k+1} = \vect f\left(\bar{\vect x}_k,\bar{\vect u}_k,\vect d_k\right),\quad \vect x_0 = \vect s \\
		&\quad \vect h\left(\bar{\vect x}_k,\bar{\vect u}_k\right) \leq 0,\quad \vect h_\mathrm{f}\left(\bar{\vect x}_N\right) \leq 0
	\end{align}
\end{subequations}
where $L\left(\bar{\vect x}_k,\bar{\vect u}_k\right)$ and $T\left(\bar{\vect x}_N\right)$ are the stage and terminal costs, respectively. $\vect s$ is the current system state while the nominal NMPC predicted trajectories are $\bar{\vect x}$ and $\bar{\vect u}$. We label $\vect h$ and $\vect h_f$ as the nominal stage and terminal inequality constraints, respectively. We will model the model error and real system stochasticity via the uncertain sequence ${\vect d}_k$ of disturbances, and we will approximate its effect on the state dynamics in a computationally effective way. More specifically, we will model the sequence ${\vect d}_k$ via an i.i.d Gaussian model with mean $\bar{\vect d}_k$ and covariance $\Lambda$ i.e.
\begin{align}
	\vect d_k\sim \mathcal N\left(\bar{\vect d}_k,\Lambda\right) \label{eq:DistGauss}
\end{align}
The first-order deviation of the model state $\vect x_k$ from its nominal trajectory $\bar{\vect x}_k$ is then given by:
\begin{align}
	\Delta \vect x_{k+1} =& \left.\frac{\partial \vect f}{\partial \vect x}\right|_{\vect x_k,\vect u_k}\Delta \vect x_{k} + \left.\frac{\partial \vect f}{\partial \vect u}\right|_{\vect x_k,\vect u_k}\Delta \vect u_{k}  \\ &\qquad + \left.\frac{\partial \vect f}{\partial \vect d}\right|_{\vect x_k,\vect u_k}\Delta \vect d_{k} \nonumber
\end{align} 
where $\Delta \vect x_0$ is a given deviation of the initial state and $\Delta \vect d_{k} = \vect d_{k} - \bar{\vect d}_k$. We observe that for $\mathbb E\left[\Delta \vect x_0\right]=0$, $\bar{\vect d}_k=\mathbb E\left[\vect d_k\right]$ and $\vect \Delta \vect u_k=0$. The expected value of the state deviation is $\mathbb E\left[ \Delta \vect x_{k}\right]=0$. We consider a linear feedback over the state deviations for the input deviation $\Delta \vect u_k$ as:
\begin{align} \label{eq:feedback}
	\Delta \vect u_{k} = -K_k\Delta \vect x_{k}
\end{align}
where $K_k$ is a control gain matrix typically based on Linear Quadratic Regulator (LQR) techniques.   
Then the state deviation dynamics read as:
\begin{align}
	\Delta \vect x_{k+1} =& \left.\left(\frac{\partial \vect f}{\partial \vect x}-\frac{\partial \vect f}{\partial \vect u}K_k\right)\right|_{\vect x_k,\vect u_k}\Delta \vect x_{k} + \left.\frac{\partial \vect f}{\partial \vect d}\right|_{\vect x_k,\vect u_k}\Delta \vect d_{k} \nonumber
\end{align} 
Let us define the time-varying matrices as follows: 
\begin{align*}
	A_k =\left. \frac{\partial \vect f}{\partial \vect x}-K_k\frac{\partial \vect f}{\partial \vect u}\right|_{\vect x_k,\vect u_k},\quad B_k = \left.\frac{\partial \vect f}{\partial \vect d}\right|_{\vect x_k,\vect u_k}
\end{align*}
Then, the covariance of the state deviation reads as:
\begin{align*}
	\Sigma_{k+1} &= \mathbb E \left[\Delta \vect x_{k+1}\Delta \vect x_{k+1}^\top  \right] \\
	&= \mathbb E \left[ \left(A_k\Delta \vect x_{k} +B_k\Delta \vect d_{k}\right)\left(A_k\Delta \vect x_{k} +B_k\Delta \vect d_{k}\right)^\top\right] \\
	&=\mathbb E \left[ A_k\Delta \vect x_{k} \Delta \vect x_{k}^\top A_k^\top +B_k\Delta \vect d_{k}\Delta \vect d_{k}^\top B_k^\top\right] \\
	&= A_k\Sigma_k A_k^\top +B_k\Lambda B_k^\top
\end{align*}
where $\Lambda$ is the covariance of the process noise. We observe that the matrices $\Sigma_k$ describe ellipsoids of state uncertainty propagations defined by the confidence regions:
\begin{align}
	\label{eq:Ellipsoid}
	R_k:=\left\{\,\Delta \vect x_k\,\left|\,\frac{1}{2}\Delta \vect x_k^\top \Sigma_k^{-1} \Delta \vect x_k \leq \sigma_k\right.\right\}
\end{align}
where $\sigma_k$ is a Mahalanobis distance (the radius of the ellipsoids as the confidence regions). More specifically, we observe that for a Gaussian disturbance $\vect d_k$ as in \eqref{eq:DistGauss} with $n$-dimensional mean vector $\bar{\vect d}_k$, the state deviation approximation $\Delta x_k$ is also Gaussian, given by the following density:
\begin{align}
	\rho\left(\Delta x_k\right) = \left(2\pi\right)^{-\frac{n}{2}}\left(\mathrm{det}\left(\Sigma_k\right)\right)^{-\frac{1}{2}}e^{-\frac{1}{2}\Delta\vect x_k^\top \Sigma_k^{-1}\Delta\vect x_k}
\end{align}
such that at time $k$, a state deviations $\Delta x_k$ has a probability density larger or equal to $ \left(2\pi\right)^{-\frac{n}{2}}\left(\mathrm{det}\left(\Sigma_k\right)\right)^{-\frac{1}{2}}e^{-\sigma_k}$ to belong to $R_k$ in \eqref{eq:Ellipsoid}.
\par 
The probability of the state deviation to belong to the ellipsoid described by \eqref{eq:Ellipsoid} is given by:
\begin{align}
	\label{eq:ProbEllips}
	\mathbb P[\Delta x_k \in R_k] = \int_{R_k} \rho\left(\Delta x_k\right) \mathrm d\Delta x_k =   \frac{\gamma_k \left(\frac{n}{2},\frac{\sigma^2_k}{2}\right) }{\Gamma_k \left(\frac{n}{2}\right) }     
\end{align}
where, $\Gamma_k $ is the Gamma function and $\gamma_k$ is the lower incomplete Gamma function describing the probability content for a n-dimensional multinormal distribution \cite{Walck}.
Imposing that the  ellipsoids described by \eqref{eq:Ellipsoid} are contained within the MPC feasible set for some sequence $\sigma_k$ then becomes a way to approximately satisfy the MPC constraints with probability \eqref{eq:ProbEllips}. In the next section, we detail how to build an inexpensive Robust MPC scheme that ensures the ellipsoids described by \eqref{eq:Ellipsoid} are contained within the MPC feasible set.
\subsection{Inclusion constraint}
Consider the constraint 
\begin{align}
	\vect h_i\left({\vect x}_k,{\vect u}_k\right) \leq 0 \label{eq:h:i}
\end{align}
for some index $i$ and suppose that we want to formulate a computationally tractable constraint requiring that \eqref{eq:h:i} is satisfied for all state deviations in the ellipsoid \eqref{eq:Ellipsoid}. For nonlinear constraints, this is unfortunately very difficult. As a result, in line with the philosophy above, we will consider the approximation resulting from linearizing the constraint on the nominal trajectory $\bar{\vect x}_k,\bar{\vect u}_k$ of the MPC scheme. More specifically, we will consider imposing the constraint:
\begin{align}
	\label{eq:Ellipsoid:Inclusion}
	\vect h_i\left(\bar{\vect x}_k,\bar {\vect u}_k\right) +&\left. \left(\frac{\partial \vect h_i}{\partial\vect x} -\frac{\partial \vect h_i}{\partial\vect u}K_k \right)\right|_{\bar{\vect x}_k,\bar {\vect u}_k}\Delta \vect x_k \leq 0 ,\\
	&\forall\,\Delta \vect x_k\in R_k \nonumber
\end{align}
Fortunately, this requirement takes a closed form. Indeed, we can simply seek the $\Delta \vect x_k\in \eqref{eq:Ellipsoid} $ that gives the worst (highest) value to $\vect h_i$.  Let us label
\begin{align*}
	\vect b_i = \left(\frac{\partial \vect h_i}{\partial\vect x} -\frac{\partial \vect h_i}{\partial\vect u}K_k  \right)^\top
\end{align*}
Then \eqref{eq:Ellipsoid:Inclusion} can be (tightly) enforced by imposing that:
\begin{align}
	\vect h_i+\sqrt{2\sigma_k}\left(\vect b_i^\top \Sigma_k\vect b_i\right)^{\frac{1}{2}} \leq 0
\end{align}
which is a constraint mixing $\vect h$ and its derivative, the dynamics of $\Sigma_k$ and the given $\sigma_k$. 
\subsection{Robust MPC formulation}
We can now formulate an approximate robust NMPC scheme:
\begin{subequations} 
	\label{eq:RNMPC}
	\begin{align}
		V_\theta(\vect s)=&\min_{\bar{\vect x},\bar{\vect u},\vect \zeta,\Sigma}\quad \gamma^N\left(T_\theta\left(\bar{\vect x}_N\right)+\vect w_{f}^\top\vect\zeta_{N}\right)\nonumber\\&
		 + \sum_{k=0}^{N-1} \gamma^k\left(l_\theta\left(\bar{\vect x}_k,\bar{\vect u}_k\right)+\vect w^\top\vect\zeta_{k}\right) \label{eq:V} \\
		\mathrm{s.t.}&\quad \bar{\vect x}_{k+1} = \vect f_\theta\left(\bar{\vect x}_k,\bar{\vect u}_k, \vect d_k\right),\quad \bar{\vect x}_0 = \vect s 	\label{eq:dyn}\\
		&\quad\Sigma_{k+1} = A_k\Sigma_k A_k^\top +B_k\Lambda B_k^\top, \quad \Sigma_0 = S_0 \label{eq:cov} \\ 
		&\quad \vect h_i\left(\bar{\vect x}_k,\bar{\vect u}_k\right)+\sqrt{2\sigma_k}\left.\left(\vect b_i^\top \Sigma_k\vect b_i\right)^{\frac{1}{2}}\right|_{\bar{\vect x}_k,\bar{\vect u}_k}\leq \vect\zeta_k\\
		&\quad\vect h_i^\mathrm{f}\left(\bar{\vect x}_N\right)+\sqrt{2\sigma_N}\left.\left(\vect b_i^\top \Sigma_N\vect b_i\right)^{\frac{1}{2}}\right|_{\bar{\vect x}_N} \leq \vect \zeta_N \\
		&\quad \vect\zeta_{0,\ldots,N} \geq 0	\label{eq:ineq}
	\end{align}
\end{subequations}
where, $0< \gamma\leq 1$ is a discount factor and $T_\theta$ is an adjustable terminal cost function. The slack variables and initial guess for the covariance matrix of the state deviation are labeled by $\vect \zeta$ and $S_0$, respectively. Although we initialize $S_0$ at zero in this paper, One can use an observer scheme to estimate this initial matrix.
The adjustable stage cost function in the above RNMPC scheme is defined as follows: 
\begin{align}
			&l_\theta\left(\bar{\vect x}_k,\bar{\vect u}_k\right)= L\left(\bar{\vect x}_k,\bar{\vect u}_k\right)+\varphi_\theta\left(\bar{\vect x}_k,\bar{\vect u}_k, \Sigma_k\right),
\end{align}
where, $\varphi_\theta\left(\bar{\vect x}_k,\bar{\vect u}_k, \Sigma_k\right)=\mathrm{Tr}\left(\frac{\partial ^2L\left(\bar{\vect x}_k,\bar{\vect u}_k\right)}{\partial \bar{\vect x}^2_k}M\Sigma_k\right)$ is proposed to adopt as a cost modification term in the stage cost $L\left(\bar{\vect x}_k,\bar{\vect u}_k\right)$ of the RNMPC scheme. This term is considered to deliver the impact of the uncertainty on the adopted stage cost. We propose to use a quadratic form for this term, which includes the adjustable matrix $M$ as an RL parameter. $\vect f_\theta$ is the nonlinear dynamics, which could be adjusted by RL through the model bias parameters. Because the robust NMPC scheme \eqref{eq:RNMPC} uses approximations and a possibly inaccurate model of the system dynamics and disturbances, it may become infeasible. To address that issue, an $\ell_1$ relaxation of the inequality constraints is introduced in \eqref{eq:ineq}. If the weights $\vect w , \vect w_{f}$ are chosen large enough, then the solution of the RNMPC scheme \eqref{eq:RNMPC} respects the constraints when it is feasible to do so.

\par 
We ought to specify here that the proposed approximate RNMPC may be interpreted either as a genuine approximate robust NMPC scheme or an approximate stochastic NMPC scheme depending on the nature of the real system disturbances. Indeed, even if the real system does not yield any approximation in \eqref{eq:RNMPC}, i.e. the disturbances are Gaussian and the system dynamics and constraints are linear, then this scheme ensures the constraint satisfaction in the sense of the probability \eqref{eq:ProbEllips}, and shall therefore be seen as a stochastic NMPC scheme. If the disturbances are bounded and properly covered by the ellipsoidal model used in \eqref{eq:RNMPC}, then the proposed scheme can be seen as a robust NMPC scheme. In both cases, the choice of parameters in \eqref{eq:RNMPC} to model the ellipsoids so as to achieve the given control objectives is difficult.
\par 
The NMPC parameters $\theta=\left \{ M, \bar {\vect d}_k,\Lambda, \sigma_k\right \}$ will be adjusted using LSTDQ-learning. We propose to use \eqref{eq:RNMPC} as an approximator for the true value function $V_\star$, i.e. we will seek the robust MPC parameters $\vect\theta$ that best achieve $V_{\vect\theta}\approx V_\star$. Let us define the control policy as:
\begin{gather}\label{eq:Policy}
	\pi_{\theta}(\vect s)=\bar{\vect u}_0^\star
\end{gather}
where, $\bar{\vect u}_0^\star$ is the first element of the input sequence $\bar{\vect u}_0^\star,\cdots,\bar{\vect u}_{N-1}^\star$ solution of \eqref{eq:RNMPC}. We next consider this optimal policy delivered by the RNMPC scheme as an action $\vect a$ in the context of reinforcement learning where it is selected according to the above policy with the possible addition of exploratory moves \cite{sutton}.
As a result in \cite{gros2019data}, a parameterized NMPC scheme can be used as an (action)-value function approximator in the context of the reinforcement learning. Then, we propose to use the parameterized RNMPC scheme \eqref{eq:RNMPC} to deliver the value function needed in the proposed second-order LSTDQ-learning algorithm. Then, the action-value function in \eqref{eq:Q} results from solving the same RNMPC scheme with its first input constrained to the delivered action $\vect a$.
\begin{subequations}\label{eq:Q}
	\begin{align}
		Q_{\theta}(\vect s,\vect a)=\min_{\bar{\vect x},\bar{\vect u},\vect \zeta,\Sigma}&\quad \eqref{eq:V} \\
		\mathrm{s.t.}&\quad \eqref{eq:dyn}-\eqref{eq:ineq}\\
		&\quad \bar{\vect u}_0=\vect a \label{eq:act}
	\end{align}
\end{subequations}
\section{RL-based Robust NMPC} \label{sec:2}

In this section, we propose to use RL to tune the approximate robust NMPC scheme. Indeed, the proposed robust NMPC scheme is based on coarse approximations of the state uncertainty propagation, which are likely to yield suboptimality and even constraints violations in practice. Finding the parameters that yield the best closed-loop performance for the real system in terms of cost and constraints violations is a difficult problem. We propose to use RL to find these parameters. More specifically, RL is used to tune the sequences of $\sigma_k$ and the expected value of the noise $\bar{\vect d}_k$, the process noise $\Lambda$ and the matrix $M$ in the cost modification $\varphi_\theta$ in order to improve the Robust NMPC scheme. RL allows one to perform this tuning based on observed state transition, and without any knowledge of the real system stochasticity.

In this section, we first present the algorithmic details needed to implement a second order LSTDQ-learning algorithm on the ellipsoidal RNMPC scheme. Then the sensitivity analysis needed in the reinforcement learning scheme is described.
\subsection{Second-Order LSTDQ Learning}
LSTDQ-based reinforcement learning methods make an efficient use of data and tend to converge faster than more basic temporal-difference learning methods. Let us form the least squares of Bellman residual error w.r.t $\theta$ as:
\begin{align}\label{eq:main_lstd}
		\min_{\theta}\mathbb{E} \left[ \| Q_\star(\vect s,\vect a)-{Q}_\theta(\vect s,\vect a) \|^2\right]  
\end{align}
where, ${Q}_\theta$ obtained from the RNMPC scheme in \eqref{eq:Q} is an approximation of the true action-value function $Q_\star$.
Let us consider the following approximation of the Bellman optimality equation, which is used in the Temporal Difference (TD)-based learning approaches.
\begin{align}\label{eq:bell}
	Q_\star(\vect s,\vect a)\approx L(\vect s,\vect a)+\gamma Q_\theta(\vect s_+,\pi(\vect s_+))  
\end{align}
where, $L(\vect s,\vect a)$ is the baseline cost in the RL scheme. Substituting \eqref{eq:bell} into \eqref{eq:main_lstd}, the least square problem \eqref{eq:main_lstd} for the first-order LSTDQ-learning can be solved for some data acquired from the transitions $\vect s_i,\vect a_i  \rightarrow \vect s_{i+1}$ as:
\begin{subequations}\label{eq:LSTDQ1}
\begin{align}
	&\mathbb E\left[ \delta \nabla_{\theta}Q_\theta(\vect s_i,\vect a_i)\right]=0,\\&
	\delta=L(\vect s_i,\vect a_i)+\gamma V_\theta(\vect s_{i+1})-Q_\theta(\vect s_i,\vect a_i)
\end{align}
\end{subequations}
where, $\delta$ is the temporal difference error. Both value and action-value functions, respectively, $V_\theta(\vect s_{i+1})=Q_\theta(\vect s_{i+1},\pi(\vect s_{i+1}))$ and $Q_\theta(\vect s_i,\vect a_i)$ are obtained from the RNMPC schemes detailed in the last section. In this paper, we adopt a Newton method to solve \eqref{eq:LSTDQ1} and extract the newton step for the second-order LSTDQ scheme as follows:
\begin{subequations}\label{eq:LSTDQ2}
	\begin{align}
         &\theta\leftarrow\theta-\alpha A^{-1}b,\\
         &A =\mathbb E\left[ \delta \nabla^2_{\theta}Q_\theta + \nabla_{\theta}Q_\theta  \nabla_{\theta}^\top \delta  \right], \quad b=\mathbb E\left[\delta\frac{\partial Q_\theta}{\partial \theta}\right]
	\end{align}
\end{subequations}
where, the scalar $\alpha>0$ is the step size.
\subsection{Sensitivity Analysis}
The gradient and Hessian of the function $Q_\theta$ needed in \eqref{eq:LSTDQ2} require one to compute the sensitivities of the optimal value of NLP \eqref{eq:Q}.
Let us define the Lagrange function $\mathcal{L}_\theta$ associated to the RNMPC problem \eqref{eq:Q} as follows:
\begin{gather}
	\mathcal{L}_{\theta}=\Phi_{\theta}+\vect\lambda^\top G_{\theta}+\vect\mu^\top H_{\theta}
\end{gather}
where $H_{\theta}$ gathers the inequality constraints of \eqref{eq:Q} and $\Phi_\theta$ is the cost of the RNMPC optimization problem.  Variable $\vect\lambda$ is the Lagrange multiplier vector associated to the equality constraints $G_\theta$ of the RNMPC. Variable $\vect\mu$ is the Lagrange multiplier vector associated to the inequality constraints of the RNMPC scheme. Let us label the primal variables for the RNMPC scheme as $\vect p=\left\{\vect X,\vect U\right\}$, where $\vect X,\vect U$ are the state and control input trajectories predicted by \eqref{eq:Q}. Then, the primal-dual variables read as $\vect z=\left\{\vect p,\vect\lambda,\vect\mu\right\}$.

The sensitivity analysis of optimization problems detailed in \cite{sensAnalysis} delivers the Gradient $\nabla_{\theta}Q_\theta$ and Hessian $\nabla^2_{\theta}Q_\theta$ ($H(Q_\theta)$) terms needed in \eqref{eq:LSTDQ2} as follows:	
\begin{subequations}\label{sens_mpc}
\begin{align}
	&\frac{\partial Q_\theta}{\partial \theta}=\frac{\partial \mathcal{L}_\theta(\vect s,\vect a,\vect z^\star)}{\partial \theta},\\
	&H(Q_\theta)=\frac{D}{D\theta} \left(\frac{\partial \mathcal{L}_\theta(\vect s, \vect a,\vect z^\star)}{\partial \theta}\right)\approx \frac{\partial^2 \mathcal{L}_\theta(\vect s,\vect a,\vect z^\star)}{\partial \theta^2} 
\end{align}
\end{subequations}		
where $\vect z^\star$ is the primal-dual solution of \eqref{eq:Q} and $D$ is the total derivative. The Hessian term $H(Q_\theta)$ can improve the search direction using the curvature information and thus make more progress per step in comparison with the
first-order LSTDQ learning relying only on the gradient calculation.
\subsection{Constrained RL steps}
In the proposed RL-RNMPC scheme, there is a matrix $\Lambda$ in the dynamics \eqref{eq:cov}, which is adjusted using the reinforcement learning method. As a requirement, this  disturbance covariance matrix must be positive semidefinite. However, the RL steps delivered by second-order LSTDQ learning do not necessarily respect this requirement, and we need to enforce it via constraints on the RL steps throughout the learning process. To address this requirement, we formulate a Semi-Definite Program (SDP) as a least squares optimization problem:
\begin{subequations}\label{sdp}
	\begin{align}
		\min_{\Delta\theta}&\quad\frac{1}{2}\left \| \Delta \theta \right \|^2-F^\top\Delta\theta\\
		\mathrm{s.t.}&\quad \Lambda({\theta + \Delta \theta})\geq 0
	\end{align}
\end{subequations}
where we assume that the above matrix is linear function of $\theta$ and the gradient term $F=-\alpha A^{-1}b$ is delivered by the second-order LSTDQ-learning algorithm \eqref{eq:LSTDQ2}.
\section{Numerical Example}\label{sec:3}
To illustrate proposed RL-RNMPC, we consider a simulated Wheeled Mobile Robot (WMR) tracking trajectories while avoiding static obstacles avoidance, and affected by model uncertainties and disturbances. Let us define the WMR model as:
\begin{align}\label{WMR_c}
	&\vect f(\vect x, \vect u)=\begin{bmatrix}
		\cos(\psi) &0 \\ 
		\sin(\psi)& 0\\ 
		0&1 
	\end{bmatrix}\vect u
\end{align}
where $\vect x=\left[x,y,\psi\right]^\top$ and $\vect u=\left[v,\omega\right]^\top$ are the state and control input vectors, respectively. The position coordinates of the WMR are labeled $x, y$ and $\psi$ is the robot orientation angle. The initial position of the mobile robot is $\vect x_0=\left[-1,2,0\right]^\top$. The control inputs $v$ and $\omega$ are the linear and angular velocities, respectively. To discretize the above continuous model, we adopt a fourth-order Runge-Kutta (RK4) integrator providing discretized function $\vect f_d$ of the WMR model. We will consider that the real system evolves according to the dynamics:
\begin{align}\label{WMR_d}
	\vect x(k+1)=\vect f_d\left(\vect x(k),\vect u(k)+\vect\Gamma_1 (k)\right)+\vect\Gamma_2(k)
\end{align}
where the two variables $\vect\Gamma_1, \vect\Gamma_2$ in the above equations model the uncertainties  as:
\begin{align}\label{diss}
	\vect \Gamma_1(k)=v(k)\begin{bmatrix}
		d_1(k)\\ 
		d_2(k)\\ 
	\end{bmatrix},\quad \vect\Gamma_2(k)=T_sv(k)\begin{bmatrix}
		0\\ 
		0\\
		d_2(k) 
	\end{bmatrix}
\end{align}
where $d_1(k)\sim \mathcal N\left(0,\Sigma_1^2\right)$ and $d_2(k)\sim \mathcal N\left(0,\Sigma_2^2\right)$. The sampling time is $T_s$= 0.2 $s$ and disturbance variances $\Sigma_1$ and $\Sigma_2$ are set as 0.2 and 0.4, respectively. We initialize the adjustable process noise covariance matrix ($3\times3$) as $\Lambda=diag(\Sigma_1^2,\Sigma_2^2,\Sigma_2^2)$. The expected value of the process noise $\bar d$ and the matrix $M$ as an RL parameter in the cost modification term $\varphi_\theta$ are initialized at zero. According to \eqref{eq:ProbEllips}, the probability of the uncertainties in the ellipsoids can be obtained from the Gamma functions. Since the simulated mobile robot has three states, there is a 3-dimensional multinormal distribution $n=3$. To ensure a large enough probability of the uncertainty propagation in the ellipsoids (confidence regions), we set $\sigma_k=2.65$. Therefore, with pair $n=3,\quad \sigma_k=2.65$ we will have the probability of the state deviations to belong to the ellipsoids $\mathbb P[\Delta x_k \in R_k]\approx 93\%$.
\par 
The RMPC model will be based on the real system dynamics, assuming that we know them perfectly. More specifically, the nominal model used in the RNMPC scheme is based on the expected value of the exact distribution of the disturbances $d_1,d_2$, and the structure according to which the disturbances enter the system is assumed known. We further observe that by assuming the disturbances are Gaussian, we are considering an ideal setup for the approximate robust MPC to perform well. Hence any gain of performance achieved by RL in this simulation setup is done through handling the approximations introduced in the robust MPC scheme. As it is proposed to adopt the RL approach in order to adjust the RNMPC parameters, we formulate a second-order LSTDQ scheme with a batch size equal to 100 data samples and let RL to update the parameters based on the collected data in the batch. In this simulation, the number of transitions (iterations) is 8000 and thus the number of RL steps is 80. The step size is set as $\alpha=10^{-6}$. We adopt a baseline stage cost used in the LSTDQ scheme as:
\begin{align}\label{eq:BL}
	L(\vect x_k,\vect u_k)=l(\vect x_k,\vect u_k)+\vect w^\top \max(0,\vect h(\vect x_k))
\end{align}
where $l(\vect x_k,\vect u_k)$ can be expressed as a quadratic function of the state and action deviations from their desired values. The second term in the above baseline is considered to cope with the violations, where $\vect h\geq 0$ is pure inequality vector of constraints induced by the obstacles. The penalty weights are $\vect w=[30,30,30]$.

\begin{figure}[htbp!]
	\centering
	\includegraphics[width=.9\linewidth]{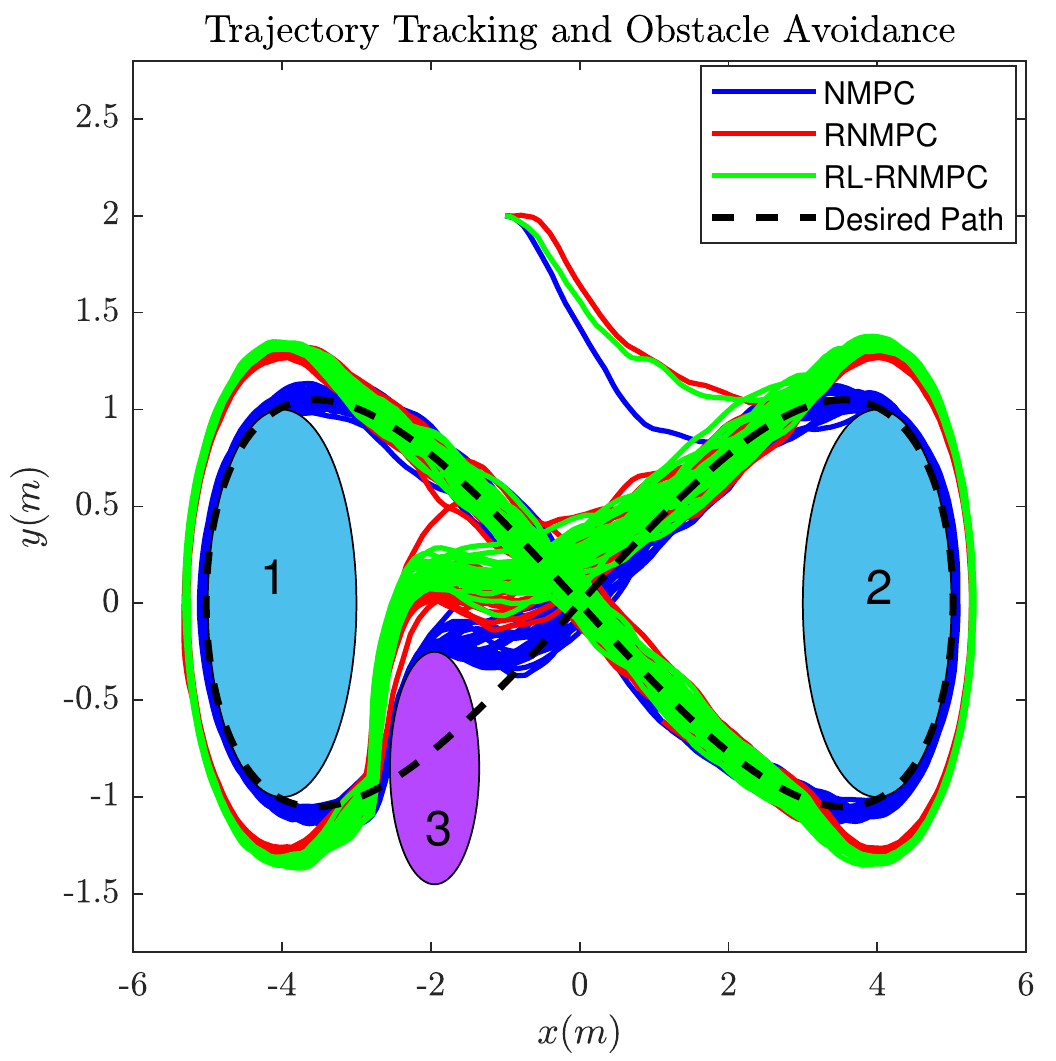}
	\caption{A comparative study: Trajectory tracking and obstacle avoidance for a WMR under uncertainties}
	\label{traj}
\end{figure}

The obstacles are represented as ellipsoids (see Fig. \ref{traj}). The reference trajectory shown in Fig.~\ref{traj} is an eight-shaped path, which intersects the obstacles or comes very close to them. 
\par
\begin{figure}[htbp!]
	\centering
	\includegraphics[width=.85\linewidth]{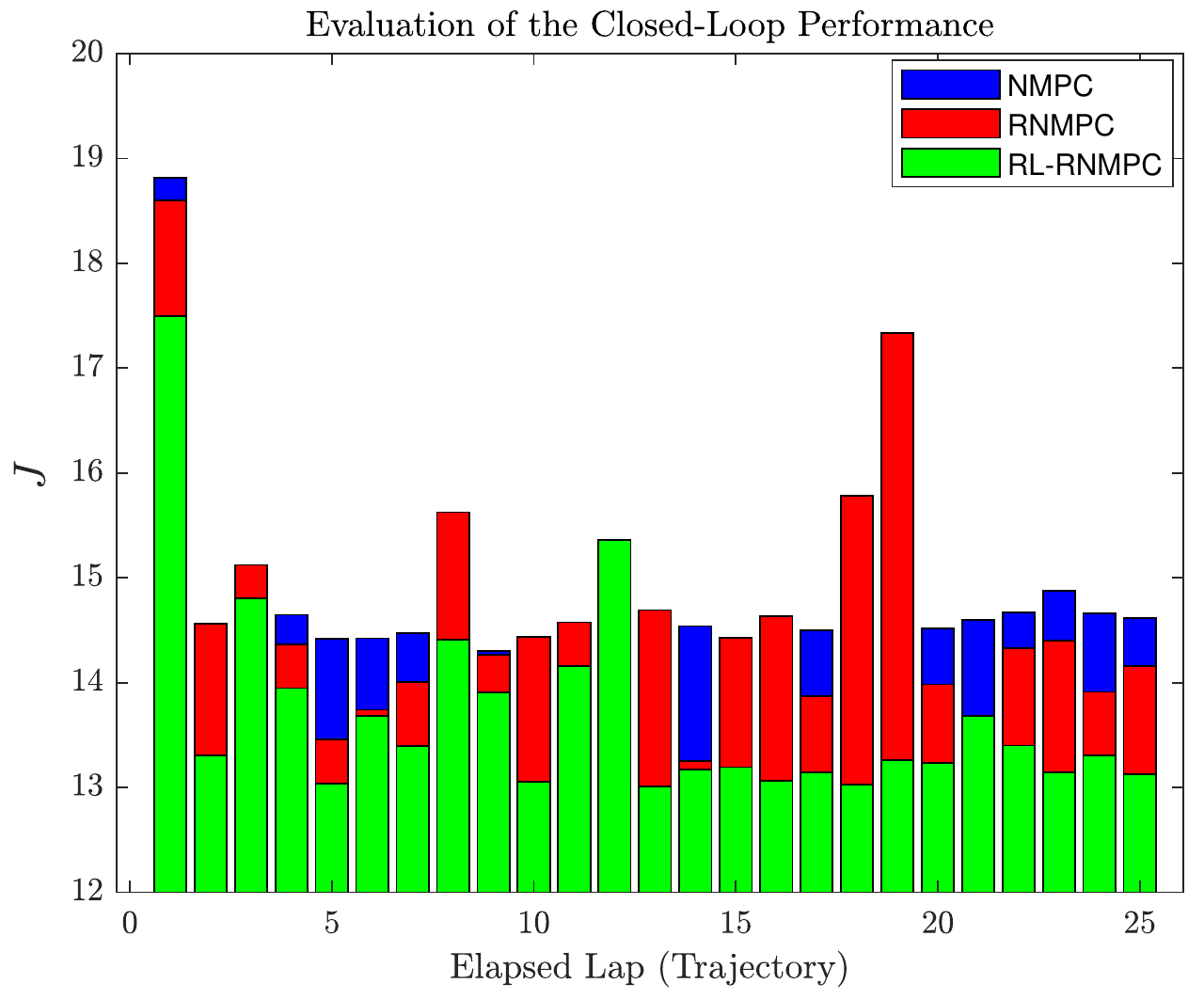}
	\caption{Average Closed-Loop performance index for 25 elapsed trajectories (laps).}
	\label{J}
\end{figure}

 
In Fig.~\ref{J}, we compare the closed-loop performance of the nominal NMPC scheme \eqref{eq:ClassicMPC}, the approximate robust NMPC without learning, and with learning. The second-order LSTDQ-learning algorithm improves the performance of the robust NMPC scheme in comparison with the classic NMPC and RNMPC without learning.

\par
\begin{figure}[htbp!]
	\centering
	\includegraphics[width=.85\linewidth]{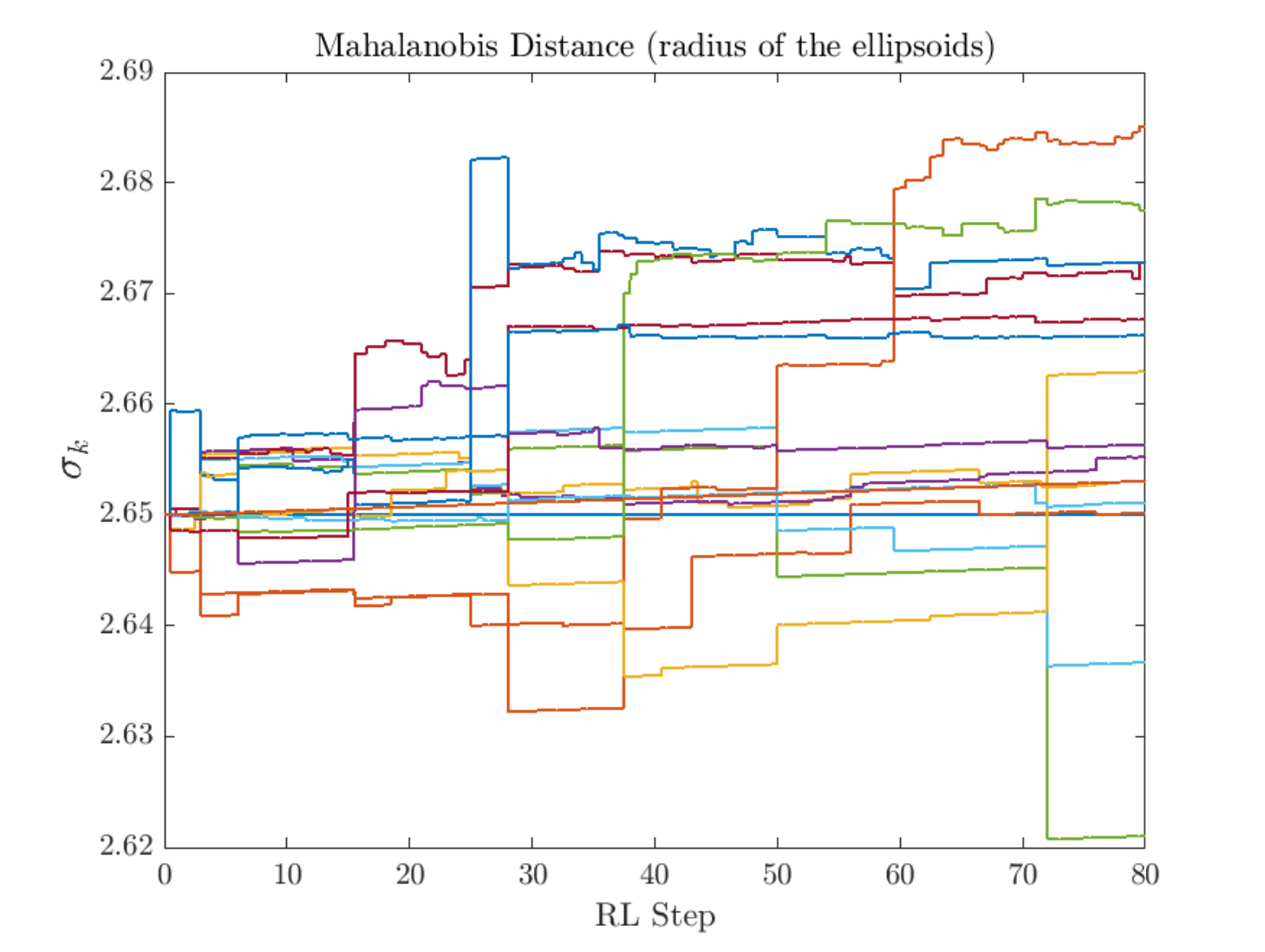}
	\caption{Tuning the ellipsoidal confidence region (adjustment of the radius) }
	\label{sig_k}
\end{figure}

 
Since the approximate RNMPC scheme is built based on an exact knowledge of the disturbance statistics and structure, and system model, the improvement of performance observed in Fig.~\ref{J} results purely on improving the approximation performed in the RNMPC scheme via modifying the robust NMPC parameters $\theta=\left \{ M, \bar {\vect d}_k,\Lambda, \sigma_k\right \}$. For majority of the ellipsoids, the dimension of the ellipsoid-shaped confidence regions $R_k$ is changed by RL to be modelled as larger than the originally selected $\sigma_k$ in order to reduce the risk of constraints violation, see Fig.~\ref{sig_k}. For some situations that the risk of hitting is not high, this dimension is decreasing using the RL. Indeed, there is a trade-off made by the RL between the avoidance and tracking accuracy in this proposed RL-RNMPC.
\begin{figure}[htbp!]
	\centering
	\includegraphics[width=.85\linewidth]{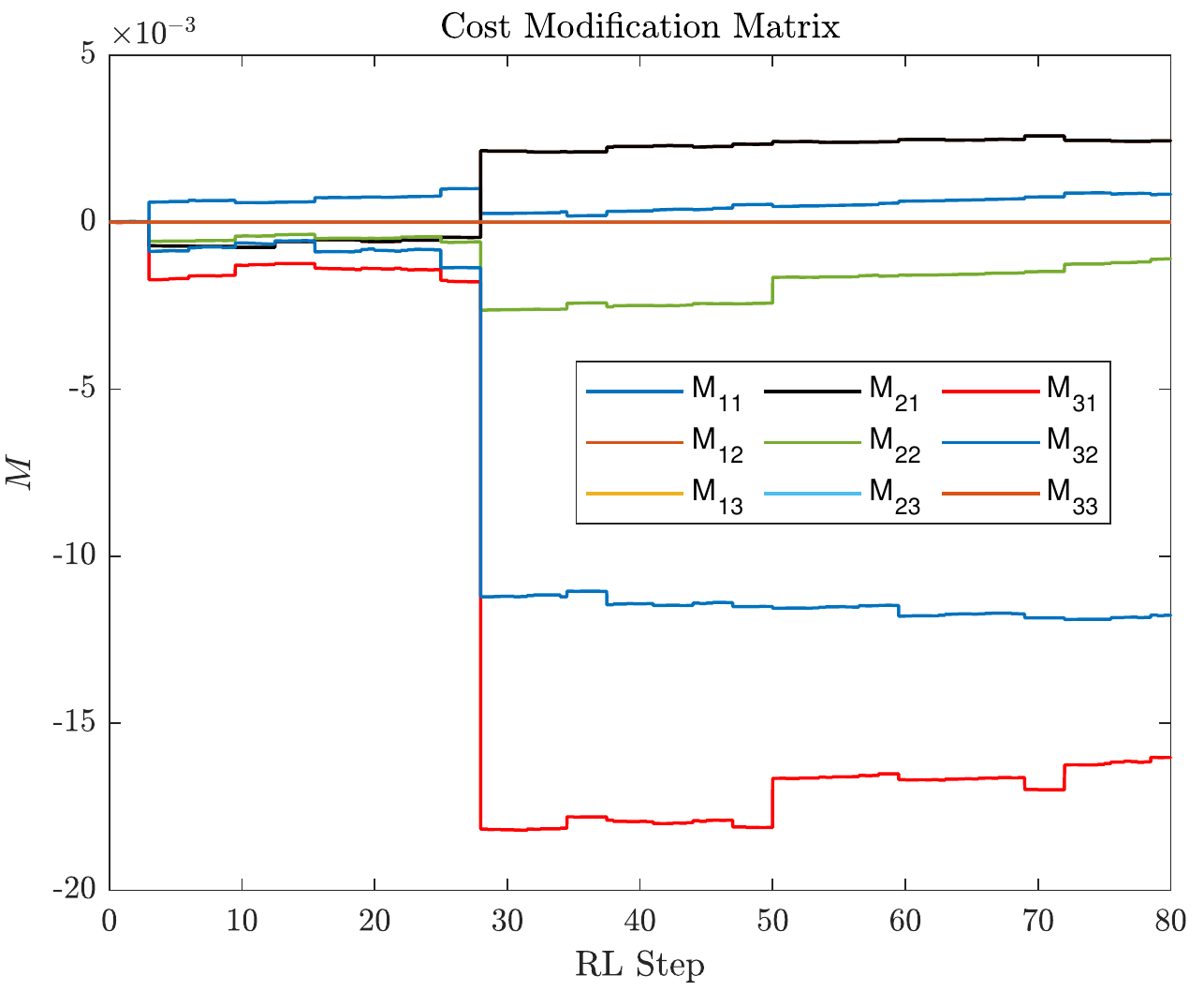}
	\caption{Matrix $M$ is adjusted as a RL parameter $\theta$ in the cost modification term $\varphi_\theta\left(\bar{\vect x}_k,\bar{\vect u}_k, \Sigma_k\right)$.}
	\label{M_cost}
\end{figure}
\begin{figure}[htbp!]
	\centering
	\includegraphics[width=.85\linewidth]{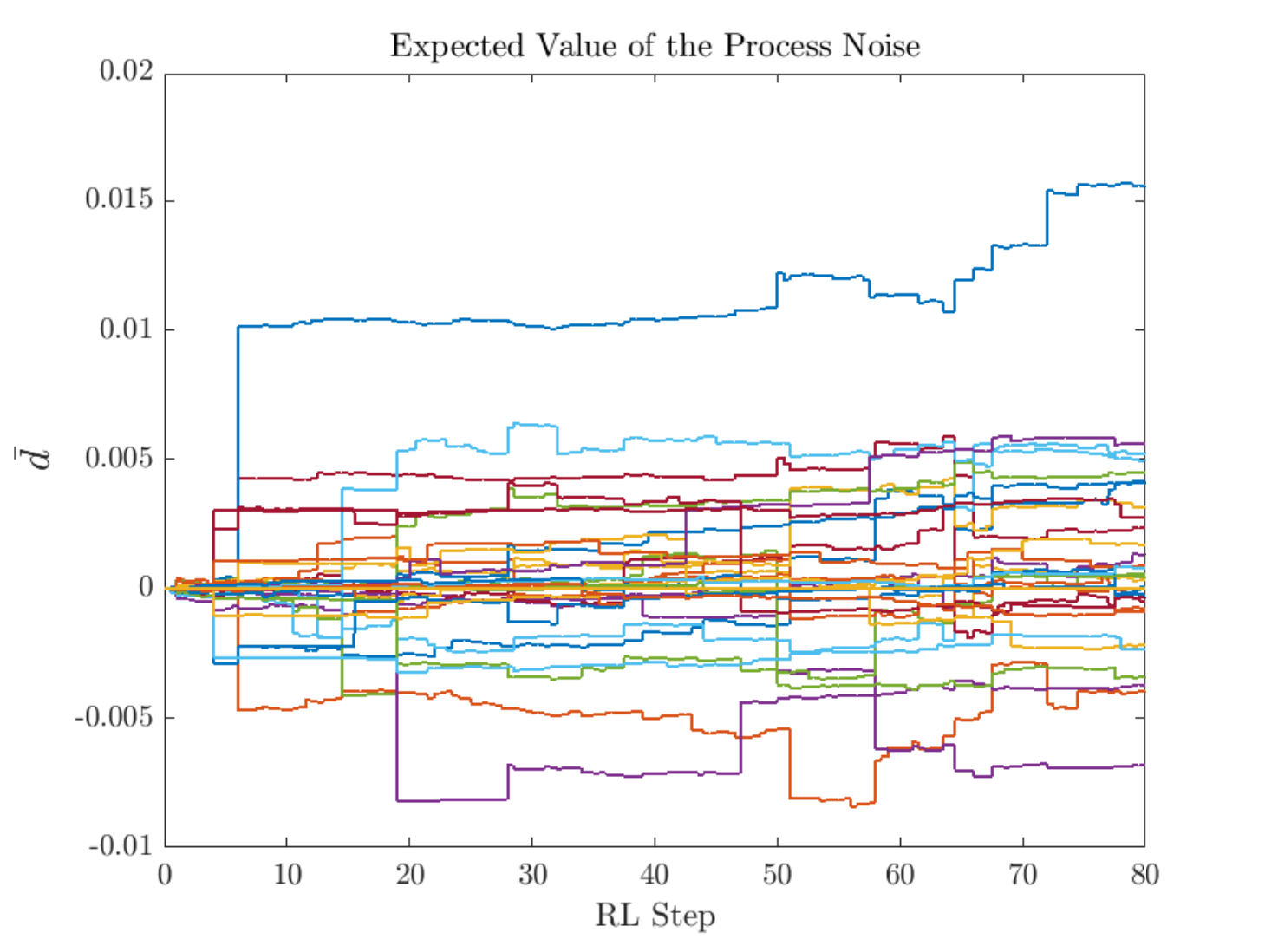}
	\caption{Tuning the expected values of the process noises applied to the three states for a prediction horizon $N=15$.}
	\label{dbar}
\end{figure}
The adjustment of the RNMPC can also contribute the mobile robot to avoid hitting the obstacles 1,2 due to possible drifting (as an uncertainty) in the road bend shown in Fig.~\ref{traj}.  The RL scheme additionally modifies the uncertainty model via $\bar d$ and $\Lambda$, see Fig.~\ref{dbar}, in order to gain performance.
\section{Conclusion}\label{sec:4}
In the context of the Robust Nonlinear Model Predictive Control (RNMPC), the formal RNMPC techniques are difficult to implement on nonlinear systems, and thus it is common to use approximate RNMPC methods instead. In this paper we proposed to describe the propagation of the uncertainties using the ellipsoidal tubes in which the disturbances and state deviations are modeled as a Gaussian noise. However, the approximated models and constraints used in this kind of RNMPC can affect the closed-loop performance and thus we adopted an LSTDQ learning as a fast RL algorithm to adjust some crucial parameters of the approximate RNMPC. In this paper, we used the proposed RNMPC as a value function approximator for the LSTDQ algorithm. AS a future work, we will propose to embed a Linear Quadratic Regulator (LQR) in the proposed RNMPC in order to deliver the first guess of the control gain matrix for the adopted linear feedback over the state deviation. Furthermore, we will investigate the proposed RL-RNMPC for a large-scale application.   

\bibliographystyle{IEEEtran}
\bibliography{IEEEabrv,Refe1}
\end{document}